\documentclass[aps,amsmath,amssymb,twocolumn]{revtex4-1}

\usepackage{graphicx}
\usepackage{dcolumn}
\usepackage{bm}
\usepackage{color}

\begin{document}

\title{Anisotropic thermal magnetoresistance for an active control of radiative heat transfer}

\author{R. M. Abraham Ekeroth$^{1,2}$}
\author{P. Ben-Abdallah$^{3,4}$}
\author{J. C. Cuevas$^{5,6}$}
\author{A. Garc\'{\i}a-Mart\'{\i}n$^1$}
\email{a.garcia.martin@csic.es}

\affiliation{$^1$IMN-Instituto de Micro y Nanotecnolog\'{\i}a (CNM-CSIC), Isaac Newton 8,
PTM, Tres Cantos, E-28760 Madrid, Spain}
\affiliation{$^{2}$Instituto de F\'{\i}sica Arroyo Seco, Universidad Nacional del Centro 
de la Provincia de Buenos Aires, Pinto 399, 7000 Tandil, Argentina}
\affiliation{$^3$Laboratoire Charles Fabry, UMR 8501, Institut d'Optique, CNRS,
Universit\'e Paris-Saclay, 2 Avenue Augustin Fresnel, 91127 Palaiseau Cedex, France}
\affiliation{$^4$Universit\'e de Sherbrooke, Department of Mechanical Engineering, Sherbrooke, 
PQ J1K 2R1, Canada}
\affiliation{$^5$Departamento de F\'{\i}sica Te\'orica de la Materia Condensada and Condensed Matter 
Physics Center (IFIMAC), Universidad Aut\'onoma de Madrid, E-28049 Madrid, Spain}
\affiliation{$^6$Department of Physics, University of Konstanz, D-78457 Konstanz, Germany}

\date{\today}

\begin{abstract}
We predict a huge anisotropic thermal magnetoresistance (ATMR) in the near-field radiative heat 
transfer between magneto-optical particles when the direction of an external magnetic field is 
changed with respect to the heat current direction. We illustrate this effect with the case of 
two InSb spherical particles where we find that the ATMR amplitude can reach values of up to 
800\% for a magnetic field of 5 T, which is many orders of magnitude larger than its spintronic 
analogue in electronic devices. This thermomagnetic effect could find broad applications in the 
fields of ultrafast thermal management as well as magnetic and thermal remote sensing.

\end{abstract}

\maketitle

In 1857 Thomson discovered that the resistivity of bulk ferromagnetic metals depends on the relative 
angle, $\theta$, between the electric current and the magnetization direction \cite{Thomson1857}. 
This phenomenon, known as anisotropic magnetoresistance (AMR), plays nowadays a central role
in the field of spintronics \cite{Zutic2004} and it is the basis of sensors for magnetic recording 
\cite{McGuire1975,Handley2000}. The AMR originates from the anisotropy of electron scattering
due to the spin-orbit interaction \cite{Handley2000}. In bulk samples the rotation of the magnetization 
leads to a relative change in the resistance that varies as $\cos^2 \theta$ with an amplitude on the 
order of 1\%, while it has been recently shown that this amplitude can be increased by an order 
of magnitude in atomic-scale ferromagnetic junctions \cite{Bolotin2006,Viret2006,Sokolov2007,Strigl2015,
Schoeneberg2016,Rakhmilevitch2016}. In this Letter we predict the existence of a thermal analogue of 
AMR in the context of radiative heat transfer between magneto-optical (MO) particles. In particular, 
we predict that, in the near-field regime, this effect can be several orders of magnitude larger than its 
spintronic counterpart.

\begin{figure}[b]
\begin{center} \includegraphics[width=0.6\columnwidth,clip]{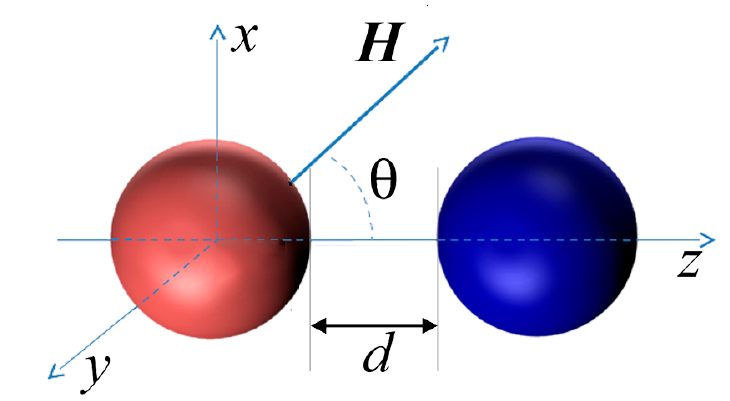} \end{center}
\caption{(Color online) Sketch of the system with two identical magneto-optical particles of radius 
$r$ held at different temperatures and separated by a gap $d$. A magnetic field $\boldsymbol H$ 
lying in the $xz$-plane is applied in a direction forming an angle $\theta$ with the heat transport 
direction ($z$-axis).}.
\label{fig-system}
\end{figure}

In recent years, the field of thermal radiation has received  a new impetus from the confirmation 
that the radiative heat transfer between two closely placed objects can greatly overcome the far-field 
limit set by the Stephan-Boltzmann law \cite{Polder1971,Kittel2005,Rousseau2009,Shen2009,Ottens2011,
Kralik2012,Zwol2012,Song2015,Kim2015,St-Gelais2016,Song2016,Bernardi2016,Cui2017,Kittel2017}. This 
enhanced thermal radiation stems from the contribution of evanescent waves (photon tunneling) that 
dominate the near-field regime. At present, one of the central challenges in this field is to actively 
control the near-field radiative heat transfer (NFRHT). In this context, magneto-optical (MO) objects 
has been put forward as a promising avenue to control the NFRHT with an external magnetic field 
\cite{Moncada-Villa2015}. In the last two years, several thermomagnetic effects have been predicted 
such as a near-field thermal Hall effect \cite{Ben-Abdallah2016}, the existence of a persistent heat 
current \cite{Zhu2016}, or a giant thermal magnetoresistance \cite{Latella2017}. However, most of the 
attention has been devoted to the role of the magnitude of the field. In this work, we show that the 
NFRHT between two MO particles can be efficiently controlled by changing the direction of the magnetic 
field, in the spirit of the AMR in spintronics. This phenomenon, which we term \emph{anisotropic thermal 
magnetoresistance} (ATMR), stems from the anisotropy of the photon tunneling induced by the magnetic field. 
We discuss this effect through the analysis of the radiative heat exchange between two InSb particles, see 
Fig.~\ref{fig-system}, and show that the ATMR can reach amplitudes of 100\% for fields on the order of 
1 T and up to 1000\% for a magnetic field of 6 T. These values are several orders of magnitude larger 
than in standard spintronic devices \cite{McGuire1975,Handley2000}.

For illustrative purposes, we consider here two identical spherical MO particles of radius $r$ embedded 
in vacuum and separated by a gap $d$, as sketched in Fig.~\ref{fig-system}. We assume that these particles 
are held at temperatures $T+\Delta T$ and $T$ with $\Delta T \ll T$ (linear response regime) and they are 
subjected to a magnetic field $\boldsymbol H$ that forms an angle $\theta$ with the axis linking the two
particles, see Fig.~\ref{fig-system}. To describe the radiative heat transfer between these two 
particles within the framework of fluctuational electrodynamics \cite{Rytov1989}, we use the thermal 
discrete dipole approximation (TDDA) of Ref.~[\onlinecite{Martin2017}]. This method allows us to compute 
numerically the thermal radiative properties of MO particles of arbitrary size and shape by discretizing 
the objects in terms of point dipoles in the spirit of the DDA approach \cite{Draine1994,Yurkin2007}. Within 
the TDDA approach, the radiative thermal conductance between two objects is given by the following Landauer-like 
formula 
\begin{equation}
G(H,\theta) = \int^{\infty}_0 \frac{d\omega}{2\pi} \frac{\partial \Theta(\omega,T)}{\partial T}
{\cal T}(\omega,H,\theta) ,
\label{eq-G}
\end{equation}
where $\Theta(\omega, T) = \hbar \omega/[\exp(\hbar \omega/k_{\rm B} T) -1]$ is the mean energy of
an harmonic oscillator in thermal equilibrium at temperature $T$ and ${\cal T}(\omega,H,\theta)$ 
is the transmission coefficient that depends on the frequency $\omega$, the magnitude of the field
$H$ and its direction. In general, this transmission coefficient has to be computed numerically 
with the TDDA approach and we refer to Ref.~[\onlinecite{Martin2017}] for the technical details. 
To get some analytical insight, we shall also make use of the dipolar approximation in which, when
the particles are small in comparison with the thermal wavelength, $\lambda_{\rm Th} = \hbar c/(k_{\rm B} T)$, 
they can be considered as single point dipoles. In this case, the transmission coefficient appearing
in Eq.~(\ref{eq-G}) adopts the form \cite{Martin2017}
\begin{equation}
{\cal T}(\omega,H,\theta) = 4 k^4_0 \mbox{Tr} \left\{\hat C \hat \chi \hat C^{\dagger} \hat \chi \right\} ,
\end{equation}
where the susceptibility tensor $\hat \chi$ and $\hat C$ are $(3 \times 3)$ matrices given by
\begin{eqnarray}
\label{eq-chi}
\hat \chi & = & \frac{1}{2i} \left( \hat \alpha - \hat \alpha^{\dagger} \right)
- \frac{k^3_0}{6\pi} \hat \alpha^{\dagger} \hat \alpha , \\
\hat C & = & \left[\hat 1 - k^4_0 \hat {\cal G} \hat \alpha \hat {\cal G} \hat \alpha \right]^{-1} 
\hat {\cal G} .
\label{eq-C}
\end{eqnarray}
Here, $k_0 = \omega/c$ and $\hat \alpha$ is the polarizability tensor of the particles that is given by 
$\hat \alpha = [\hat \alpha_0^{-1} - i \hat 1 k^3_0/(6\pi)] ^{-1}$, where $\hat \alpha_{0} =  
3V (\hat \epsilon - \hat 1) (\hat \epsilon + 2 \hat 1)^{-1}$ is the quasistatic polarizability tensor
\cite{deSousa2016}. Here, $V = (4/3) \pi r^3$ is the volumen of the particles and $\hat \epsilon$ is the 
corresponding permittivity tensor. Finally, $\hat {\cal G}$ is the dyadic Green tensor given by \cite{Novotny2012}
\begin{eqnarray}
\hat {\cal G}(\mathbf r_1, \mathbf r_2) & = & \frac{e^{ik_0 \rho}}{4\pi \rho} \left[ \left( 1 +
\frac{ik_0 \rho -1}{k^2_0 \rho^2} \right) \hat 1 + \right. \nonumber \\ & & \left. 
\left( \frac{3 - 3ik_0 \rho -k^2_0 \rho^2}{k^2_0 \rho^2} \right) 
\frac{\boldsymbol \rho \otimes \boldsymbol \rho}{\rho^2} \right] ,
\label{eq-GF}
\end{eqnarray}
where $\mathbf r_{1,2}$ are the positions of the point dipoles, $\boldsymbol \rho = \mathbf r_1 - 
\mathbf r_2$, $\rho = |\mathbf r_1 - \mathbf r_2|$, and $\otimes$ denotes the exterior product. 

As an example of a MO material we consider $n$-doped InSb, a polar semiconductor, that
when subjected to an external magnetic field becomes MO. For an a magnetic field lying
on the $xz$ plane and forming an angle $\theta$ with the $z$-axis, see Fig.~\ref{fig-system},
the permittivity tensor of InSb adopts the form \cite{Palik1976}
\begin{equation}
\label{eq-perm-tensor}
\hat \epsilon = \left( \begin{array}{ccc} \epsilon_1 \cos^2\theta + \epsilon_3 \sin^2 \theta  
& -i\epsilon_2 \cos \theta & \frac{1}{2} (\epsilon_1 - \epsilon_3) \sin(2\theta) \\
i \epsilon_2 \cos \theta & \epsilon_1 & i\epsilon_2 \sin \theta \\ 
\frac{1}{2} (\epsilon_1 - \epsilon_3) \sin(2\theta) & -i\epsilon_2 \sin \theta & 
\epsilon_1 \sin^2 \theta + \epsilon_3 \cos^2 \theta \end{array} \right) ,
\end{equation}
where
\begin{eqnarray}
\epsilon_1(H) & = & \epsilon_{\infty} \left( 1 + \frac{\omega^2_L - \omega^2_T}{\omega^2_T - 
\omega^2 - i \Gamma \omega} + \frac{\omega^2_p (\omega + i \gamma)}{\omega [\omega^2_c -
(\omega + i \gamma)^2]} \right) , \nonumber \\
\label{eq-epsilons}
\epsilon_2(H) & = & \frac{\epsilon_{\infty} \omega^2_p \omega_c}{\omega [(\omega + i \gamma)^2 -
\omega^2_c]} , \\
\epsilon_3 & = & \epsilon_{\infty} \left( 1 + \frac{\omega^2_L - \omega^2_T}{\omega^2_T -
\omega^2 - i \Gamma \omega} - \frac{\omega^2_p}{\omega (\omega + i \gamma)} \right) . \nonumber
\end{eqnarray}
Here, $\epsilon_{\infty}$ is the high-frequency dielectric constant, $\omega_L$ is the longitudinal
optical phonon frequency, $\omega_T$ is the transverse optical phonon frequency, $\omega^2_p =
ne^2/(m^{\ast} \epsilon_0 \epsilon_{\infty})$ defines the plasma frequency of free carriers of density 
$n$ and effective mass $m^{\ast}$, $\Gamma$ is the phonon damping constant, and $\gamma$ is the 
free-carrier damping constant. Finally, the magnetic field enters in these expressions via the cyclotron 
frequency $\omega_c = eH/m^{\ast}$. In what follows, we shall consider the following set of parameters
taken from Ref.~[\onlinecite{Palik1976}]: $\epsilon_{\infty} = 15.7$, $\omega_L = 3.62 \times 10^{13}$ 
rad/s, $\omega_T = 3.39\times 10^{13}$ rad/s, $\Gamma = 5.65 \times 10^{11}$ rad/s, $\gamma = 3.39 \times 10^{12}$ 
rad/s, $n = 1.07 \times 10^{17}$ cm$^{-3}$, $m^{\ast}/m = 0.022$, and $\omega_p = 3.14 \times 10^{13}$ rad/s.

\begin{figure}[t]
\begin{center} \includegraphics[width=0.95\columnwidth,clip]{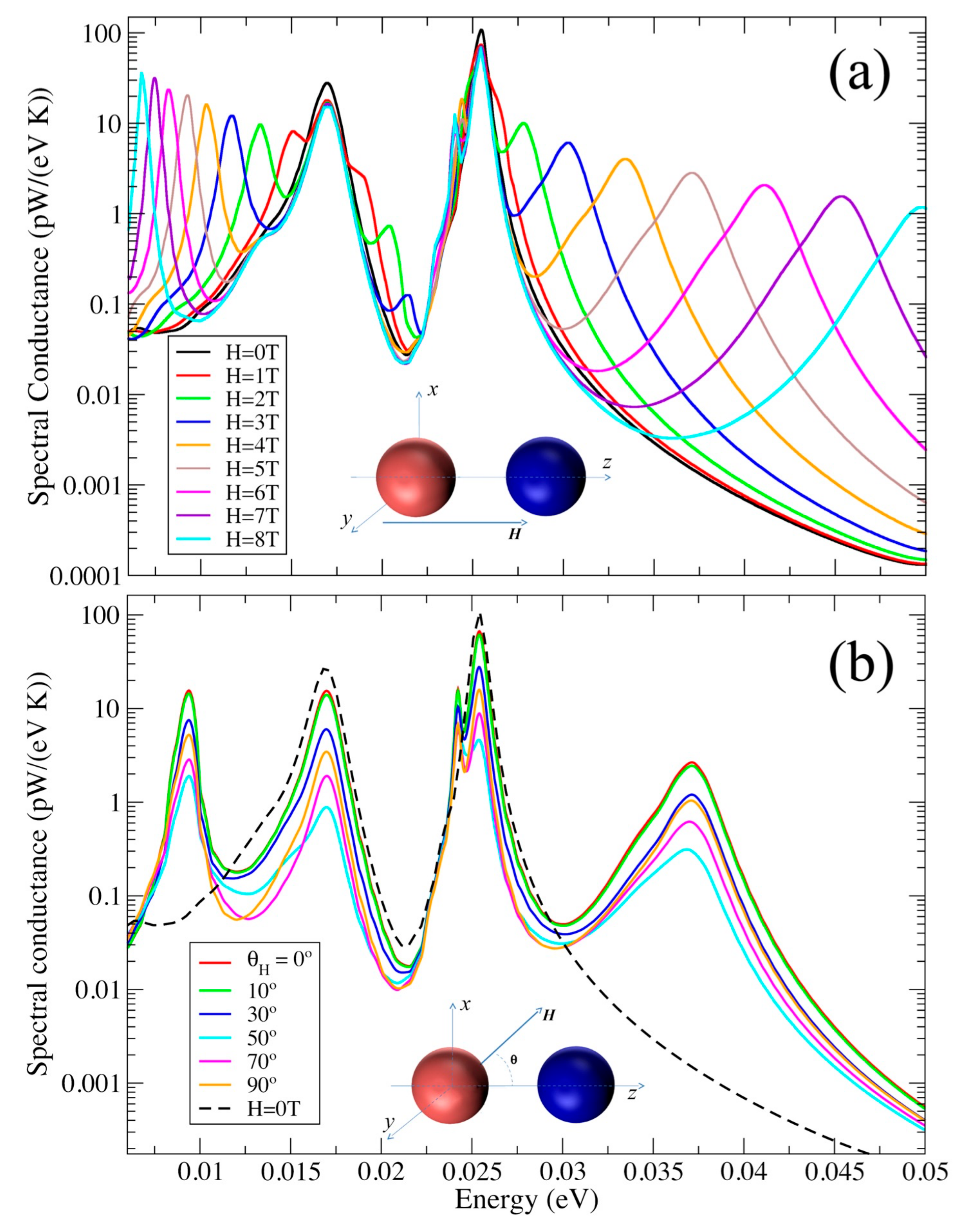} \end{center}
\caption{(Color online) (a) Room temperature spectral conductance as a function of the photon energy 
for two InSb spheres of radius 250 nm separated by a gap of 500 nm for different values of the 
magnetic field point along the $z$-direction, see inset. (b) The corresponding spectral conductance 
for a magnetic field magnitude of 5 T and different angles between the magnetic field and the transport
direction, see inset. The dashed line correspond to the case for zero field.}
\label{fig-spectral-RHT}
\end{figure}
\begin{figure}[t]
\begin{center} \includegraphics[width=0.85\columnwidth,clip]{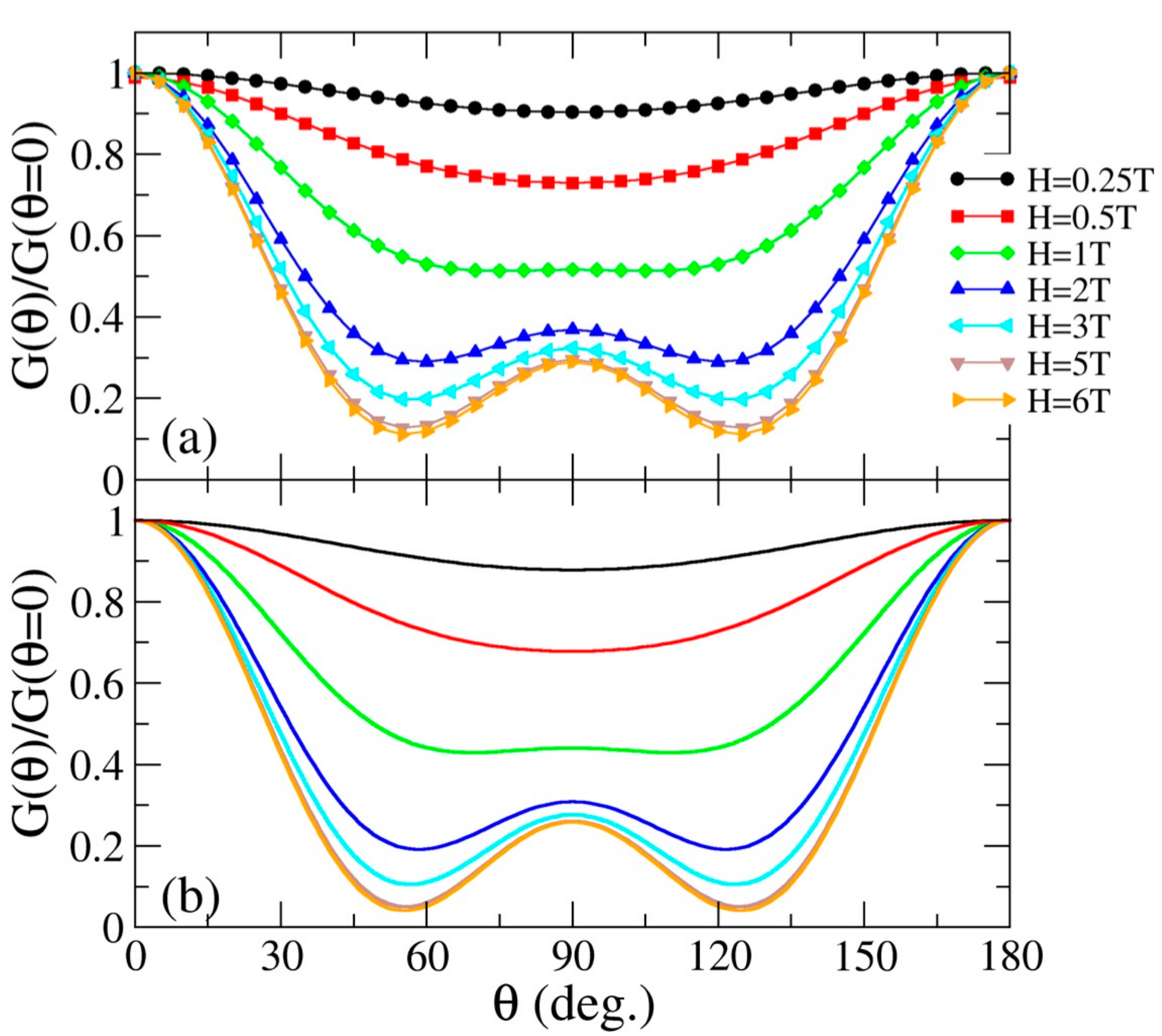} \end{center}
\caption{(Color online) (a) Room-temperature thermal conductance as a function of the angle $\theta$ 
between the magnetic field and the transport direction for two InSb particles of radius 250 nm, a gap 
of 500 nm and different values of the magnetic field magnitude. The conductance is normalized by the 
conductance at $\theta=0$. (b) The same as in panel (a), but calculated with the dipolar approximation.}
\label{fig-ATMR}
\end{figure}

Let us start the discussion of the results by analyzing the effect of the magnitude of the magnetic
field in the radiative heat transfer. In Fig.~\ref{fig-spectral-RHT}(a) we show the spectral conductance,
defined as conductance per unit of photon energy, for two InSb particles of radius 250 nm and a gap of 
500 nm. The different curves correspond to different values of the magnetic field that is directed along 
the transport direction ($z$-axis). These results were computed with the TDDA method of Ref.~\cite{Martin2017} 
discretizing each particle in 1791 cubic dipoles, which was checked to be enough to converge the results. As 
seen in Fig.~\ref{fig-spectral-RHT}(a), the spectral conductance in the absence of field is dominated by two 
peaks that, as we shall show below, are related to the localized plasmons of these particles. As the 
field increases, new peaks appear that disperse with the field following the magnetic-field-induced 
localized plasmons of these particles (see below). Notice that in some energy regions the magnetic field 
has a dramatic effect and it changes the spectral conductance by more than two orders of magnitude. Let us
remark that the spectral conductance and the transmission coefficient, see Eq.~(\ref{eq-G}), have basically 
the same energy dependence because at room temperature thermal factor $\partial \Theta(\omega,T)/ \partial T$
is almost constant in the energy region of interest. 

Let us now explore the role of the field direction. For this purpose, we show in Fig.~\ref{fig-spectral-RHT}(b) 
the dependence of the spectral conductance on the angle between the magnetic field and the transport 
direction for a fixed field magnitude of $H = 5$ T. As one can see, the rotation of the field strongly 
modulates the height of the spectral conductance peaks, but it does not change their position in energy. 
This modulation is the essence of the ATMR effect and originates from the 
magnetic-field-induced anisotropy in the photon tunneling. To quantify this effect we define the ATMR ratio 
$G(\theta)/G(\theta=0)$ between the conductance at a given angle $\theta$ and the conductance when the field 
points along the transport direction ($\theta=0$). The numerical results obtained with TDDA for this ATMR 
ratio for the example of Fig.~\ref{fig-spectral-RHT} are shown in Fig.~\ref{fig-ATMR}(a) for different 
values of the field amplitude. As one can see, the conductance is strongly modulated by the field direction 
and it is symmetric around $\theta = 90^{\rm o}$. Notice also that, irrespective of the field value, the 
conductance is maximum at $\theta = 0$. Moreover, for low fields the conductance reaches a minimum 
at $\theta = 90^{\rm o}$, while at higher fields (above 1 T) the conductance exhibits two minima away from 
$90^{\rm o}$. More importantly, the ATMR ratio reaches, e.g., a minimum of 0.51 for $H = 1$ T and of 0.127 
for $H=5$ T. In terms of a thermal resistance, $R=1/G$, these ATMR ratios imply relative changes 
$[R(\theta)-R(\theta=0)]/R(\theta=0)$ of approximately 95\% and 700\%, respectively, which are truly 
remarkable when we compare them with the 1\% relative change in the resistance of spintronic devices 
for similar fields \cite{Handley2000}.  

\begin{figure}[t]
\begin{center} \includegraphics[width=0.9\columnwidth,clip]{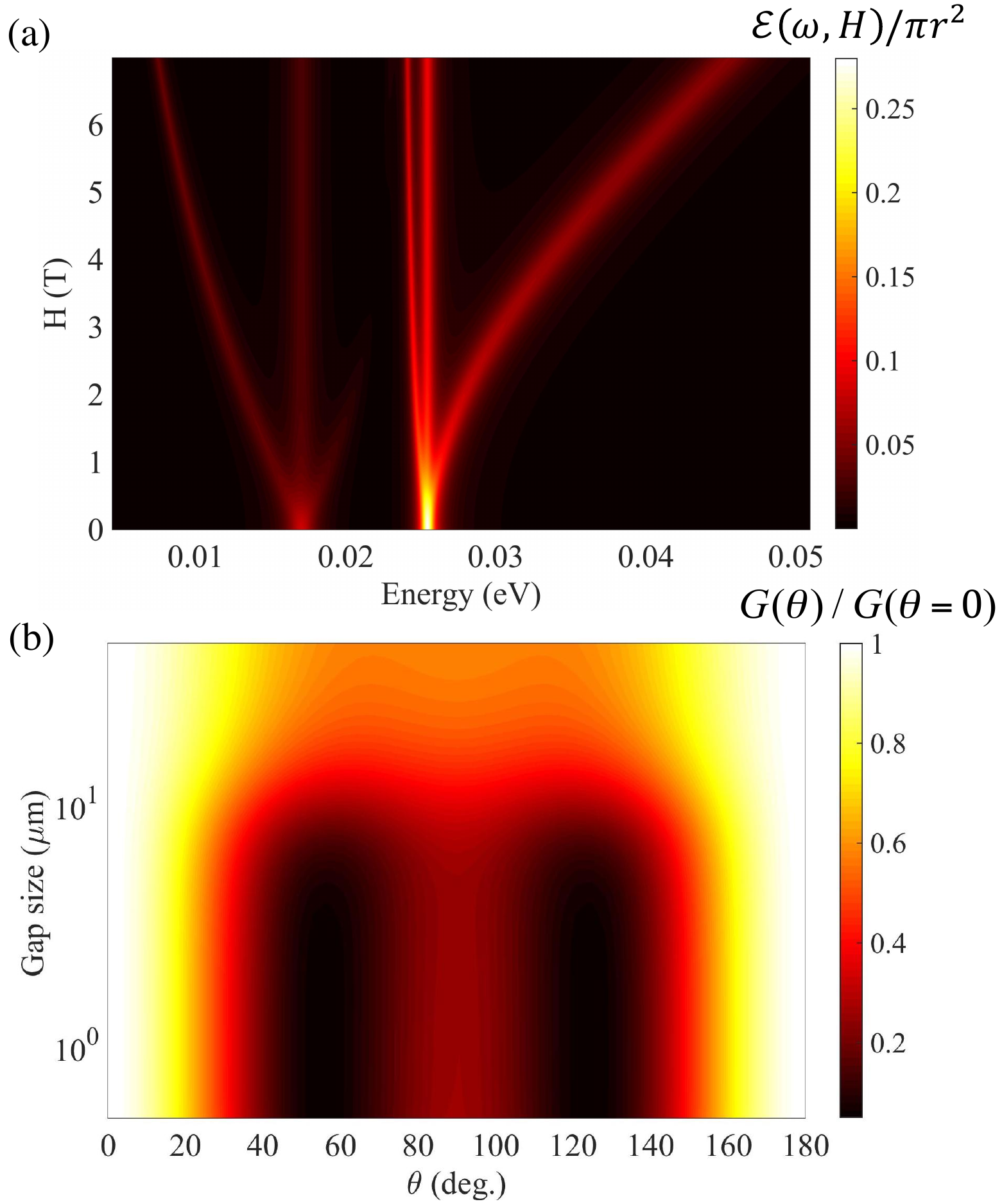} \end{center}
\caption{(Color online) (a) Total spectral emissivity of a InSb dipole of radius 250 nm as a 
function of the photon energy and the magnitude of the magnetic field. The emissivity is normalized
by the geometrical cross section $\pi r^2$. (b) Room-temperature thermal conductance as a function 
of the angle $\theta$ between the magnetic field and the transport direction and as a 
function of the gap for two InSb spheres of radius 250 nm and a field magnitude $H = 5$ T. The 
conductance is normalized by the conductance at $\theta=0$ for every gap and the calculations were 
done using the dipolar approximation.}
\label{fig-modes}
\end{figure}

In order to understand these results, we now resort to the dipolar approximation in which each particle is 
considered to be a single dipole. Using Eqs.~(\ref{eq-G}-\ref{eq-GF}) we have computed the ATMR for the same 
parameters as in Fig.~\ref{fig-ATMR}(a) and the results are shown in panel (b) of that figure. As one can see, the 
dipolar approximation qualitatively reproduces all the salient features discussed in the previous paragraph
and this, in turn, justifies the use of this approximation to elucidate the underlying physics. As a next step,
we want to understand the origin of the peaks in the spectral conductance, see Fig.~\ref{fig-spectral-RHT}(a).
For this purpose, we investigate the electromagnetic modes supported by a single dipole. A convenient way
to reveal the frequency and field dependence of these modes is to plot the total spectral emissivity of a 
single dipole, which in turn is equal to the total absorption cross section (its sum over the three spatial 
directions). This emissivity is given by \cite{Martin2017} ${\cal E}(\omega,H) = (1/3)k_0 \mbox{Tr}\{\hat \chi\}$ 
and it is shown in Fig.~\ref{fig-modes}(a) as a function of the photon energy and the magnetic field amplitude 
for a particle radius of 250 nm. Notice that the maxima of the emissivity nicely correspond to the peaks in the 
spectral conductance in Fig.~\ref{fig-spectral-RHT}(a). These maxima reveal the existence of two electromagnetic 
modes at zero field, which become up to six at finite field. In fact, from an analysis of the quasistatic polarizability, 
$\hat \alpha_0$, one can show that these resonant modes can be also obtained from the solutions of
\begin{equation} 
\det \left( \hat \epsilon(\omega,\boldsymbol H) + 2 \hat 1 \right) = 0 = (\epsilon_3 + 2)
\left[ (\epsilon_1 + 2)^2 - \epsilon^2_2 \right] .
\label{eq-modes}
\end{equation}
At zero field, $\epsilon_1 = \epsilon_3 = \epsilon$ and $\epsilon_2 = 0$, and the previous condition 
reduces to the well-known condition for localized plasmons \cite{Bohren1998}: $\epsilon(\omega)=-2$. 
Notice that the condition of Eq.~(\ref{eq-modes}) is independent of $\theta$, as expected
since we are dealing with an overall property of a spherical particle. This explains why the peak
positions in the spectral conductance in Fig.~\ref{fig-spectral-RHT}(b) do not depend on the field
direction. 

Now, to understand the angular dependence of the ATMR, we use the fact that in the relevant 
frequency range the polarizability can be approximated by the quasistatic polarizability ($\hat \alpha 
\approx \hat \alpha_0$). With this idea, and neglecting multiple scattering, i.e., using $\hat C \approx
\hat {\cal G}$ in Eq.~(\ref{eq-C}), it is straightforward to show that the ATMR ratio in the dipolar 
approximation adopts the form 
\begin{equation}
\frac{G(\theta)}{G(\theta=0)} = 1 + A \sin^2 \theta + B \sin^4 \theta ,
\label{eq-G-theta}
\end{equation}
where $A < 0$ and $B>0$ (with $B < |A|$) are two coefficients that depend on the magnetic field amplitude 
and the gap. Moreover, one can show that at low fields (below 0.1 T), the relative change in the conductance
goes as $[G(\theta)-G(\theta=0)]/G(\theta=0) \propto -H^2 \sin^2(\theta)$, i.e., it is quadratic with 
the field. Equation~(\ref{eq-G-theta}) describes very accurately the angular dependence shown in 
Fig.~\ref{fig-ATMR}, for both the exact results and those obtained with the dipolar approximation. This
also explains why at low fields there is a single minimum at $\theta = 90^{\rm o}$ and two minima for higher 
fields. Physically, this angular dependence arises from the anisotropic thermal emission of these particles 
induced by the external magnetic field. This anisotropy leads in the near-field regime to the corresponding 
anisotropy in the photon tunneling.  

The strong modulation of the ATMR is quite generic and it appears in a wide range of parameters. 
We illustrate this fact in Fig.~\ref{fig-modes}(b) where we show the ATMR ratio computed within the 
dipolar approximation as a function of both $\theta$ and the gap size ranging from 500 nm to 50 $\mu$m 
for particles of radius 250 nm. As one can see, the angular dependence of the ATMR ratio remains rather 
constant in the near-field regime, i.e., for gaps below 10 $\mu$m, and its amplitude diminishes in the 
far-field regime. In all cases, this dependence is accurately  described by Eq.~(\ref{eq-G-theta}). 

Let us now discuss some potential applications of the ATMR effect. For small changes in the field direction, 
the heat conductance varies quadratically with the angle $\theta$, which shows that the heat flow can be 
efficiently modulated with a small variation of the magnetic field. The operational speed of such a modulation 
is only limited by the induction mechanisms used to tune the magnetic field so that the ATMR effect could be 
used to investigate heat transfer at few picoseconds scale. Thus, it could be used to investigate 
out-of-equilibrium heat transfer processes at the time scale of heat carrier relaxation. In addition, the 
ATMR effect could be implemented to make local measurements of temperature gradient as well as a thermal 
sensing of magnetic field orientation.

So in summary, we have predicted a thermal analogue of the AMR that is widely used in spintronics
and proposed its use for an active control of near-field radiative heat transfer between MO objects. 
We have illustrated this effect with the case of two InSb particles and shown that the amplitude of the
modulation of the magnetoresistance can be orders of magnitude larger than in electronic devices. 
This effect paves the way for an ultrafast thermal management with external magnetic fields.

We acknowledge funding from the Spanish MINECO (FIS2014-53488-P and MAT2014-58860-P) and the 
Comunidad de Madrid (S2013/MIT-2740). P.B.-A.\ acknowledges funding support from the Discovery 
Grant Program of CRSNG and J.C.C.\ thanks the DFG and SFB767 for sponsoring his stay 
at the University of Konstanz as Mercator Fellow.



\begin{thebibliography}{00}

\bibitem{Thomson1857}
W. Thomson, 
\emph{On the Electro-Dynamic Qualities of Metals: Effects of Magnetization on the Electric 
Conductivity of Nickel and of Iron},
Proc. R. Soc. London {\bf 8}, 546 (1857).

\bibitem{Zutic2004}
I. \v{Z}uti\'c, J. Fabian, and S. Das Sarma,
\emph{Spintronics: Fundamentals and Applications},
Rev. Mod. Phys. {\bf 76}, 323 (2004).

\bibitem{McGuire1975}
T.~R. McGuire and R.~I. Potter, 
\emph{Anisotropic Magnetoresistance in Ferromagnetic 3d Alloys},
IEEE Trans. Magn. {\bf 11}, 1018 (1975).

\bibitem{Handley2000}
R.~C. O'Handley, 
\emph{Modern Magnetic Materials: Principles and Applications} (Wiley, New York, 2000), p. 573.

\bibitem{Bolotin2006}
K.~I. Bolotin, F. Kuemmeth, and D.~C. Ralph, 
\emph{Anisotropic Magnetoresistance and Anisotropic Tunneling Magnetoresistance
due to Quantum Interference in Ferromagnetic Metal Break Junctions},
Phys. Rev. Lett. {\bf 97}, 127202 (2006).

\bibitem{Viret2006}
M. Viret, M. Gabureac, F. Ott, C. Fermon, C. Barreteau, G. Autes,  R. Guirardo-Lopez,
\emph{Giant Anisotropic Magneto-Resistance in Ferromagnetic Atomic Contacts},
Eur. Phys. J. B {\bf 51}, 1 (2006).

\bibitem{Sokolov2007}
A. Sokolov, E.~Y. Tsymbal, J. Redepenning, and B. Doudin, 
\emph{Quantized Magnetoresistance in Atomic-Size Contacts},
Nat. Nanotechnol. {\bf 2}, 171 (2007).

\bibitem{Strigl2015}
F. Strigl, C. Espy, M. B\"uckle, E. Scheer, T. Pietsch,
\emph{Emerging Magnetic Order in Platinum Atomic Contacts and Chains},
Nat. Commun. {\bf 6}, 6172 (2015).

\bibitem{Schoeneberg2016}
J. Sch\"oneberg, F. Otte, N. Ne\'el, A. Weismann, Y. Mokrousov, J. Kr\"oger, R. Berndt, and S. Heinze,
\emph{Ballistic Anisotropic Magnetoresistance of Single-Atom Contacts},
Nano Lett. {\bf 16}, 1450 (2016).

\bibitem{Rakhmilevitch2016}
D. Rakhmilevitch, S. Sarkar, O. Bitton, L. Kronik, O. Tal,
\emph{Enhanced Magnetoresistance in Molecular Junctions by Geometrical
Optimization of Spin-Selective Orbital Hybridization},
Nano Lett. {\bf 16}, 1741 (2016).

\bibitem{Polder1971}
D. Polder and M, Van Hove,
\emph{Theory of Radiative Heat Transfer Between Closely Spaced Bodies},
Phys. Rev. B {\bf 4}, 3303 (1971).

\bibitem{Kittel2005}
A. Kittel, W. M\"uller-Hirsch, J. Parisi, S.-A. Biehs, D. Reddig, and M. Holthaus,
\emph{Near-field Heat Transfer in a Scanning Thermal Microscope},
Phys. Rev. Lett. {\bf 95}, 224301 (2005).

\bibitem{Rousseau2009}
E. Rousseau, A. Siria, G. Jourdan, S. Volz, F. Comin, J. Chevrier, and J.-J, Greffet,
\emph{Radiative Heat Transfer at the Nanoscale},
Nat. Photon. {\bf 3}, 514 (2009).

\bibitem{Shen2009}
S. Shen, A. Narayanaswamy, and G. Chen, 
\emph{Surface Phonon Polaritons Mediated Energy Transfer between Nanoscale Gaps},
Nano Lett. {\bf 9}, 2909 (2009).

\bibitem{Ottens2011}
R.~S. Ottens, V. Quetschke, S. Wise, A.~A. Alemi, R. Lundock, G. Mueller, D.~H. Reitze, 
D.~B. Tanner, and B.~F. Whiting,
\emph{Near-Field Radiative Heat Transfer between Macroscopic Planar Surfaces},
Phys. Rev. Lett. {\bf 107}, 014301 (2011).

\bibitem{Kralik2012}
T. Kralik, P. Hanzelka, M. Zobac, V. Musilova, T. Fort, and M. Horak, 
\emph{Strong Near-Field Enhancement of Radiative Heat Transfer between Metallic Surfaces},
Phys. Rev. Lett. {\bf 109}, 224302 (2012).

\bibitem{Zwol2012}
P.~J. van Zwol, L. Ranno, and J. Chevrier,
\emph{Tuning Near Field Radiative Heat Flux through Surface Excitations with a Metal 
Insulator Transition},
Phys. Rev. Lett. {\bf 108}, 234301 (2012).

\bibitem{Song2015}
B. Song, Y. Ganjeh, S. Sadat, D. Thompson, A. Fiorino, V. Fern\'andez-Hurtado, 
J. Feist, F.~J. Garcia-Vidal, J.~C. Cuevas, P. Reddy, and E. Meyhofer, 
\emph{Enhancement of Near-Field Radiative Heat Transfer Using Polar Dielectric Thin Films},
Nat. Nanotechnol. {\bf 10}, 253 (2015).

\bibitem{Kim2015}
K. Kim, B. Song, V. Fern\'andez-Hurtado, W. Lee, W. Jeong, L. Cui, D. Thompson,
J. Feist, M.~T.~H. Reid,  F.~J. Garcia-Vidal, J.~C. Cuevas, E. Meyhofer, and P. Reddy, 
\emph{Radiative Heat Transfer in the Extreme Near Field},
Nature (London) {\bf 528}, 387 (2015).

\bibitem{St-Gelais2016}
R. St-Gelais, L. Zhu, S.~H. Fan, and M. Lipson,
\emph{Near-Field Radiative Heat Transfer between Parallel Structures in the Deep 
Subwavelength Regime},
Nat. Nanotechnol. {\bf 11}, 515 (2016).

\bibitem{Song2016}
B. Song, D. Thompson,	 A. Fiorino,	Y. Ganjeh, P. Reddy, E. Meyhofer,
\emph{Radiative Heat Conductances between Dielectric and Metallic Parallel 
Plates with Nanoscale Gaps},
Nat. Nanotechnol. {\bf 11}, 509 (2016).

\bibitem{Bernardi2016}
M.~P. Bernardi, D. Milovich, and M. Francoeur,
\emph{Radiative Heat Transfer Exceeding the Blackbody Limit between 
Macroscale Planar Surfaces Separated by a Nanosize Vacuum Gap},
Nat. Commun. {\bf 7}, 12900 (2016).

\bibitem{Cui2017}
L. Cui, W. Jeong, V. Fern\'andez-Hurtado, J. Feist, F.J. Garc\'{\i}a-Vidal, 
J.C. Cuevas, E. Meyhofer, and P. Reddy,
\emph{Study of Radiative Heat Transfer in {\AA}ngstr\"om- and Nanometre-sized Gaps},
Nat. Commun. {\bf 8}, 14479 (2017).

\bibitem{Kittel2017} 
K. Kloppstech, N. K\"onne, S.-A. Biehs, A. W. Rodriguez, L. Worbes, D. Hellmann, and A. Kittel, 
\emph{Giant Heat Transfer in the Crossover Regime between Conduction and Radiation}, 
Nat. Commun. {\bf 8}, 14475 (2017).

\bibitem{Moncada-Villa2015}
E. Moncada-Villa, V. Fern\'andez-Hurtado, F.~J. Garc\'{\i}a-Vidal,
A. Garc\'{\i}a-Mart\'{\i}n, and J.~C. Cuevas, 
\emph{Magnetic Field Control of Near-Field Radiative Heat Transfer and the Realization 
of Highly Tunable Hyperbolic Thermal Emitters},
Phys. Rev. B {\bf 92}, 125418 (2015).

\bibitem{Ben-Abdallah2016}
P. Ben-Abdallah, 
\emph{Photon Thermal Hall Effect},
Phys. Rev. Lett. {\bf 116}, 084301 (2016).

\bibitem{Zhu2016}
L. Zhu and S. Fan,
\emph{Persistent Directional Current at Equilibrium in Nonreciprocal Many-Body 
Near Field Electromagnetic Heat Transfer},
Phys. Rev. Lett. {\bf 117}, 134303 (2016).

\bibitem{Latella2017}
I. Latella and P. Ben-Abdallah,
\emph{Giant Thermal Magnetoresistance in Plasmonic Structures},
Phys. Rev. Lett. {\bf 118}, 173902 (2017).

\bibitem{Rytov1989}
S.~M. Rytov, Y.~A. Kravtsov, and V.~I. Tatarskii,
\emph{Principles of Statistical Radiophysics} Vol.\ 3 (Springer-Verlag, Heidelberg, 1989).

\bibitem{Martin2017}
R.~M. Abraham Ekeroth, A. Garc\'{\i}a-Mart\'{\i}n, and J.~C. Cuevas, 
\emph{Thermal Discrete Dipole Approximation for the Description of Thermal Emission and 
Radiative Heat Transfer of Magneto-Optical Systems},
Phys. Rev. B {\bf 95}, 235428 (2017).

\bibitem{Draine1994}
B.~T. Draine and P.~J. Flatau, 
\emph{Discrete-Dipole Approximation for Scattering Calculations},
J. Opt. Soc. Am. A {\bf 11}, 1491 (1994).

\bibitem{Yurkin2007}
M.~A. Yurkin and A.~G. Hoekstra, 
\emph{The Discrete Dipole Approximation for Simulation of Light Scattering by Particles much 
Larger than the Wavelength},
J. Quant. Spectrosc. Radiat. Transfer {\bf 106}, 558 (2007).

\bibitem{deSousa2016}
N. de Sousa, L.~S. Froufe-P\'erez, J.~J. S\'aenz, A. Garc\'{\i}a-Mart\'{\i}n,
\emph{Magneto-Optical Activity in High Index Dielectric Nanoantennas},
Sci. Rep. {\bf 6}, 30803 (2016).

\bibitem{Novotny2012}
L. Novotny and B. Hecht, \emph{Principles of Nano-Optics},
(Cambridge University Press, Cambridge, 2012).

\bibitem{Palik1976}
E.~D. Palik, R. Kaplan, R.~W. Gammon, H. Kaplan, R.~F. Wallis, and J.~J. Quinn,
\emph{Coupled Surface Magnetoplasmon-Optic-Phonon Polariton Modes on InSb},
Phys. Rev. B {\bf 13}, 2497 (1976).

\bibitem{Bohren1998}
C.~F. Bohren and D.~R. Huffman, 
\emph{Absorption and Scattering of Light by Small Particles} (Wiley, New York, 1998).

\end{thebibliography}
\end{document}